\documentclass[aps,prc,superscriptaddress,reprint,floatfix]{revtex4-2}
\usepackage{graphicx}
\usepackage{multirow}
\usepackage{color}
\usepackage{amssymb}
\usepackage{bbm}
\usepackage{amsmath}
\usepackage{lineno}
\usepackage{enumitem}
\usepackage{scalerel,stackengine}
\usepackage{xcolor}
\usepackage{comment}
\usepackage{booktabs}
\usepackage[normalem]{ulem}
\usepackage[colorlinks=true, breaklinks=true, linkcolor=blue, citecolor=purple, urlcolor=teal]{hyperref}
\usepackage{algorithm}
\usepackage{algpseudocode}

\newcommand{\br}{{\mathbf r}}
\newcommand{\bR}{{\mathbf R}}
\newcommand{\bk}{{\mathbf k}}

\def\nuc#1#2{\relax\ifmmode{}^{#1}{\protect\text{#2}}\else${}^{#1}$#2\fi}

\newcommand{\be}{\begin{eqnarray}}
\newcommand{\ee}{\end{eqnarray}}

\newcommand{\bwt}{\begin{widetext}}
\newcommand{\ewt}{\end{widetext}}

\begin{document}

\title{Reduced basis emulator for elastic scattering in continuum-discretized coupled-channels calculations}

\author{Jin Lei}
\email[]{jinl@tongji.edu.cn}
\affiliation{School of Physics Science and Engineering, Tongji University, Shanghai 200092, China.}

\begin{abstract}
I develop a reduced basis emulator for continuum-discretized coupled-channel (CDCC) calculations that achieves speedups of $\approx 10^2$ while maintaining sub-percent accuracy. The emulator is constructed using the proper orthogonal decomposition (POD) method applied to snapshots of CDCC solutions computed at sampled points in the optical potential parameter space. The prediction is performed via Galerkin projection onto the reduced basis. I demonstrate the method using deuteron scattering on $^{58}$Ni at 21.6 MeV as a test case, emulating 18 optical potential parameters simultaneously. The emulator reproduces elastic scattering cross sections with errors below 0.1\% across a wide parameter range. This development enables efficient uncertainty quantification and Bayesian parameter estimation for nuclear reaction calculations that were previously computationally prohibitive.
\end{abstract}


\date{\today}

\maketitle

\section{Introduction}
\label{sec:intro}

Nuclear reactions involving weakly bound projectiles occupy a central position in modern nuclear physics. The advent of radioactive ion beam (RIB) facilities has opened unprecedented opportunities to study exotic nuclei far from stability~\cite{Jensen2004,Tanihata2013,Jonson2004}. Experiments at these facilities have revealed remarkable phenomena including nuclear halos, where valence nucleons extend far beyond the dense nuclear core. Understanding the reaction mechanisms of these loosely bound systems is essential not only for extracting nuclear structure information but also for nuclear astrophysics, where capture and breakup reactions play crucial roles in stellar nucleosynthesis.

A distinctive feature of reactions with weakly bound projectiles is the strong coupling to breakup channels. Unlike reactions with tightly bound nuclei where the projectile remains intact, weakly bound systems can easily dissociate during the collision, and this breakup process significantly modifies the elastic scattering cross sections and other observables~\cite{Cubero2012,DiPietro2012,Moro2007,FernandezGarcia2013,MoroLay2012}. Properly accounting for these breakup effects requires treating the coupling between bound and continuum states of the projectile, a task that goes beyond simple perturbative approaches.

The continuum-discretized coupled-channel (CDCC) method~\cite{Austern1987,Yahiro2012} has emerged as the standard theoretical framework for this purpose. In CDCC, the projectile continuum is discretized into a finite set of momentum bins, which are then treated as effective discrete states coupled to the ground state. This converts the three-body scattering problem into a tractable coupled-channel calculation while preserving the essential physics of breakup. The method has been successfully applied to a wide range of systems, from deuteron-induced reactions~\cite{Johnson1970} to exotic halo nuclei such as $^{6}$He, $^{11}$Li, and $^{11}$Be~\cite{Moro2007,FernandezGarcia2013,Cubero2012,DiPietro2012}.

Despite its success in describing experimental data, the predictive power of CDCC calculations is limited by uncertainties in the input optical potentials that describe the interactions between projectile fragments and the target nucleus. These potentials are typically parametrized using phenomenological forms with parameters constrained by fitting to elastic scattering data~\cite{Koning2003}. However, even well-constrained optical potentials carry parameter uncertainties that propagate to the predicted reaction observables. As experimental precision continues to improve at RIB facilities, there is growing need for theoretical predictions with quantified uncertainties, a requirement for meaningful comparison between theory and experiment.

Rigorous uncertainty quantification (UQ) requires exploring the high-dimensional parameter space of optical potentials through techniques such as Bayesian inference~\cite{Dobaczewski2014,Furnstahl2020,Neufcourt2019,Ekstrom2019}. Such analyses typically require $10^4$--$10^6$ evaluations of the forward model to adequately sample the parameter space. However, a single CDCC calculation can take minutes to hours depending on the complexity of the system, making brute-force UQ computationally prohibitive with conventional solvers~\cite{Descouvemont2016}.

Emulator techniques offer a promising path forward. The basic idea is to construct a fast surrogate model that approximates the expensive full calculation while being orders of magnitude faster to evaluate. In nuclear physics, eigenvector continuation~\cite{Frame2018,Konig2020,Sarkar2021} has emerged as a powerful emulation technique for bound-state calculations, enabling rapid UQ in nuclear structure. For scattering problems, several emulator approaches have been developed based on the Kohn variational principle~\cite{Furnstahl2020b,Zhang2022}, other variational methods~\cite{Drischler2021,Melendez2021}, the R-matrix theory~\cite{Bai2021,Bai2022}, and the complex scaling method~\cite{Liu2024}. Recently, coupled-channel emulators have been proposed for systems with a few channels~\cite{CatacoraRios2025,Liao2025}. However, extending these approaches to realistic CDCC calculations, which involve tens of coupled channels and complex coupling structures, remains an open challenge.

In this work, I develop a reduced basis emulator specifically designed for CDCC calculations. The approach is based on the proper orthogonal decomposition (POD) method~\cite{Quarteroni2016,Benner2015}, which constructs an optimal low-dimensional basis from a set of full CDCC solutions computed at sampled parameter values. The key insight is that, despite the high dimensionality of the solution space, CDCC solutions for different optical potential parameters lie approximately in a low-dimensional subspace, a structure that can be exploited for efficient emulation. The prediction is performed via Galerkin projection onto the reduced basis, achieving speedups of two orders of magnitude with sub-percent accuracy. I demonstrate the method using deuteron scattering on $^{58}$Ni, emulating 18 optical potential parameters simultaneously while reducing computation time from seconds to milliseconds.

The paper is organized as follows. Section~\ref{sec:theory} reviews the CDCC formalism and the direct boundary matching method (DBMM) used for solving the coupled equations. Section~\ref{sec:emulator} describes the construction and implementation of the reduced basis emulator. Section~\ref{sec:results} presents numerical results for deuteron scattering on $^{58}$Ni, demonstrating the accuracy and efficiency of the emulator. Section~\ref{sec:conclusion} summarizes my findings and discusses future applications.

\section{Theoretical Framework}
\label{sec:theory}

\subsection{CDCC formalism}
\label{sec:cdcc}

I consider the scattering of a two-body projectile $a = b + x$ from a target nucleus $A$~\cite{Austern1987,LeiMoro2023}. The effective three-body Hamiltonian is
\begin{equation}
    H = H_{\text{proj}} + T_{\bR} + U_b + U_x,
    \label{eq:hamiltonian}
\end{equation}
where $H_{\text{proj}} = T_{\br} + V_{bx}$ is the projectile internal Hamiltonian, with $T_{\br}$ the kinetic energy operator for the relative coordinate $\br$ between the projectile constituents and $V_{bx}$ the binding potential. The operator $T_{\bR}$ is the kinetic energy for the projectile-target relative motion, and $U_b$, $U_x$ are optical potentials describing the $b$-$A$ and $x$-$A$ elastic scattering at the same energy per nucleon as the incident projectile.

From the perspective of the projectile internal system, the model space can be expressed through the unit operator:
\begin{equation}
    \mathbbm{1} = \sum_{n_0} |\phi_{n_0}\rangle\langle\phi_{n_0}| + \int d\bk \, |\phi_{\bk}^{(+)}\rangle\langle\phi_{\bk}^{(+)}|,
\end{equation}
where the sum runs over the discrete spectrum of $H_{\text{proj}}$ (bound states) and the integral covers its continuous spectrum (scattering states).

For practical calculations, the continuum is discretized into momentum bins using the bin method~\cite{Yahiro2012}. The bin-state wave functions are constructed as superpositions of scattering eigenstates:
\begin{equation}
    \phi_n^{lj}(r) = \sqrt{\frac{2}{\pi N_k}} \int_{k_n}^{k_{n+1}} g(k) f_{lj}(k,r) \, dk,
    \label{eq:bin_state}
\end{equation}
where $g(k)$ is a weight function, $N_k = \int_{k_n}^{k_{n+1}} |g(k)|^2 dk$ is the normalization constant, and $f_{lj}(k,r)$ is the radial scattering wave function at momentum $\hbar k$. This discretization converts the three-body continuum problem into a finite-dimensional coupled-channel problem.

The CDCC wave function for a given total angular momentum $J$ is expanded as:
\begin{equation}
    \Psi^{JM}(\br, \bR) = \sum_{\alpha} \frac{\chi_\alpha^J(R)}{R} \mathcal{Y}_\alpha^{JM}(\hat{R}, \br),
    \label{eq:cdcc_expansion}
\end{equation}
where $\alpha = \{n, l, j, L\}$ denotes the channel quantum numbers (internal state index $n$, internal orbital angular momentum $l$, internal total angular momentum $j$, and projectile-target orbital angular momentum $L$), $\chi_\alpha^J(R)$ is the radial wave function in channel $\alpha$, and $\mathcal{Y}_\alpha^{JM}$ contains the angular and internal structure dependence.

Projecting the Schr\"odinger equation onto the bin states yields a system of coupled radial equations:
\begin{equation}
    \left[ -\frac{\hbar^2}{2\mu} \frac{d^2}{dR^2} + \frac{L(L+1)\hbar^2}{2\mu R^2} + \epsilon_n - E \right] \chi_\alpha^J + \sum_{\alpha'} U_{\alpha\alpha'}^J \chi_{\alpha'}^J = 0,
    \label{eq:coupled_channel}
\end{equation}
where $\mu$ is the projectile-target reduced mass, $\epsilon_n = \langle\phi_n|H_{\text{proj}}|\phi_n\rangle$ is the bin energy, $E$ is the total center-of-mass energy, and $U_{\alpha\alpha'}^J(R)$ is the coupling potential matrix element computed from the fragment-target optical potentials.

\subsection{Direct boundary matching method}
\label{sec:dbmm}

I solve the coupled equations~(\ref{eq:coupled_channel}) using the direct boundary matching method (DBMM)~\cite{Lei2025} based on Lagrange-Legendre basis functions~\cite{Baye2010}. The key feature of DBMM is the decomposition of the wave function into incoming and scattered components, which directly leads to a linear system with a source term.

For an incident channel $\alpha_0$, the radial wave function in channel $\alpha$ is decomposed as:
\begin{equation}
    \chi_{\alpha}^J(R) =e^{i\sigma_{L_{\alpha_0}}} \left[ \delta_{\alpha\alpha_0}  F_{L_{\alpha_0}}(\eta_{\alpha_0}, k_{\alpha_0} R) + \chi_{\alpha}^{J,\text{sc}}(R)\right],
    \label{eq:wf_decomposition}
\end{equation}
where $F_L$ is the regular Coulomb function representing the incoming wave (present only in the entrance channel), $\sigma_L$ is the Coulomb phase shift, and $\chi_{\alpha}^{J,\text{sc}}$ is the scattered wave. Substituting into the coupled equations yields an inhomogeneous equation for the scattered wave:
\begin{align}
    &\left[ -\frac{\hbar^2}{2\mu} \frac{d^2}{dR^2} + \frac{L(L+1)\hbar^2}{2\mu R^2} + U_{\alpha\alpha} - E_\alpha \right] \chi_{\alpha}^{J,\text{sc}} \nonumber \\
    &+ \sum_{\alpha'} U_{\alpha\alpha'} \chi_{\alpha'}^{J,\text{sc}} = -U_{\alpha\alpha_0}^{\text{eff}} F_{L_{\alpha_0}},
    \label{eq:inhomogeneous}
\end{align}
For the entrance channel ($\alpha = \alpha_0$), only the short-range part of the diagonal potential contributes, $U_{\alpha\alpha_0}^{\text{eff}} = U_{\alpha_0\alpha_0}^{\text{sr}} = U_{\alpha_0\alpha_0} - Z_a Z_A e^2/R$, because the point Coulomb interaction is already accounted for in the regular Coulomb function $F_{L_{\alpha_0}}$. For other channels, the full coupling potential $U_{\alpha\alpha_0}^{\text{eff}} = U_{\alpha\alpha_0}$ (including any Coulomb multipole couplings) drives the transition. The scattered wave satisfies the outgoing boundary condition $\chi_{\alpha}^{J,\text{sc}}(R) \propto H_{L_\alpha}^+(\eta_\alpha, k_\alpha R)$ at large $R$.

The radial domain $[0, R_{\max}]$ is discretized using $N$ Lagrange-Legendre mesh points, and the scattered wave is expanded as:
\begin{equation}
    \chi_{\alpha}^{J,\text{sc}}(R) = R^{-1/2} \sum_{i=1}^N c_{\alpha,i}^J \hat{f}_i(R/R_{\max}),
\end{equation}
where $\hat{f}_i(x)$ are $x$-regularized Lagrange functions defined on the interval $[0,1]$. The outgoing wave boundary condition is directly incorporated into the last row of the matrix equation by imposing:
\begin{align}
    &\left. \frac{d\chi_{\alpha}^{J,\text{sc}}}{dR} \right|_{R_{\max}} = \gamma_{\alpha} \, \chi_{\alpha}^{J,\text{sc}}(R_{\max}), \nonumber \\
    &\gamma_\alpha = k_\alpha \frac{H_{L_\alpha}^{+\prime}(\eta_\alpha, k_\alpha R_{\max})}{H_{L_\alpha}^+(\eta_\alpha, k_\alpha R_{\max})}.
\end{align}

This yields a block matrix equation:
\begin{equation}
    \mathbf{M}^J(\boldsymbol{\theta}) \, \mathbf{c}^J = \mathbf{b}^J,
    \label{eq:linear_system}
\end{equation}
where the source vector $\mathbf{b}^J$ arises from the right-hand side of Eq.~(\ref{eq:inhomogeneous}). The system matrix decomposes as $\mathbf{M}^J(\boldsymbol{\theta}) = \mathbf{K}^J + \mathbf{V}^J(\boldsymbol{\theta})$, where $\mathbf{K}^J$ contains parameter-independent contributions (kinetic energy, centrifugal barrier, boundary conditions) and $\mathbf{V}^J(\boldsymbol{\theta})$ contains the parameter-dependent coupling potentials. This decomposition is exploited in the emulator construction.

Once the coefficients are obtained, the $S$-matrix elements are extracted:
\begin{equation}
    S_{\alpha\alpha_0}^J = \delta_{\alpha\alpha_0} + 2ik_\alpha \frac{\chi_{\alpha}^{J,\text{sc}}(R_{\max})}{H_L^+(\eta_\alpha, k_\alpha R_{\max})}.
\end{equation}

\section{Reduced Basis Emulator}
\label{sec:emulator}

\subsection{Motivation}
\label{sec:motivation}

The computational challenge in CDCC arises from two factors. First, the coupled-channel equations must be solved for each partial wave $J = 0, 1, \ldots, J_{\max}$, where $J_{\max}$ increases with the mass and charge of the colliding nuclei (reaching $J_{\max} \approx 100$--$200$ for heavy systems). Second, each partial wave involves solving a large linear system: for a typical model space with $N_{\text{ch}} \approx 30$--$50$ channels (larger calculations may involve hundreds or thousands of channels) and $N \approx 100$--$200$ radial mesh points, the system~(\ref{eq:linear_system}) has dimension $N_{\text{tot}} = N_{\text{ch}} \times N \approx 5000$--$10000$. The computational cost depends on the system complexity and hardware; a complete calculation summing over all $J$ values can take seconds to hours depending on the number of channels and partial waves involved.

This cost becomes prohibitive when performing Bayesian inference to extract optical potential parameters from experimental data. Unlike traditional $\chi^2$ fitting that finds a single best-fit parameter set, Bayesian analysis explores the full posterior probability distribution, providing rigorous uncertainty quantification and revealing correlations between parameters. This is essential for making reliable predictions with quantified uncertainties. However, sampling the posterior typically requires $10^4$--$10^6$ likelihood evaluations, each involving a complete CDCC calculation. For the optical potentials considered here (Woods-Saxon forms for both proton-target and neutron-target interactions), there are 18 parameters even without spin-orbit terms, making the parameter space high-dimensional and the sampling computationally demanding.

The emulator dramatically accelerates these calculations by exploiting a simple physical observation: as the optical potential parameters change smoothly, the scattering wave functions also change smoothly. For example, increasing the depth of the real potential by 10\% does not produce a completely different wave function---it produces a wave function that looks similar to the original, with modest modifications. This smoothness means that if we know the solutions at a few representative parameter values, we can approximate the solution at any nearby parameter value as a linear combination of these known solutions. Importantly, the solution vector $\mathbf{c}^J$ contains the wave functions for \textit{all} coupled channels at all mesh points (dimension $N_{\text{tot}} = N_{\text{ch}} \times N$), so the basis solutions capture the correlated behavior across all channels simultaneously. In practice, only a small number of such basis solutions (typically 5--50) are needed to accurately span the space of solutions across the entire parameter range of interest.

\subsection{Training: building the reduced basis}
\label{sec:training}

The emulator construction proceeds in two stages: an offline training phase (expensive but done only once) and an online prediction phase (fast, repeated many times).

First, I select $N_s$ (the number of training samples) representative parameter sets $\{\boldsymbol{\theta}_1, \boldsymbol{\theta}_2, \ldots, \boldsymbol{\theta}_{N_s}\}$ that cover the parameter space of interest. These are chosen using Latin hypercube sampling~\cite{McKay1979}, which ensures the samples are spread evenly throughout the parameter space rather than clustered in certain regions. For each parameter set, I solve the full CDCC problem to obtain the solution vector $\mathbf{c}^J(\boldsymbol{\theta}_k)$. These $N_s$ solutions are called ``snapshots'' because they capture the behavior of the system at specific parameter values.

The snapshots are arranged as columns of a matrix:
\begin{equation}
    \mathbb{C}_{\text{snap}}^J = \begin{pmatrix} \mathbf{c}^J(\boldsymbol{\theta}_1) & \mathbf{c}^J(\boldsymbol{\theta}_2) & \cdots & \mathbf{c}^J(\boldsymbol{\theta}_{N_s}) \end{pmatrix},
\end{equation}
where $\mathbb{C}_{\text{snap}}^J$ has dimensions $N_{\text{tot}} \times N_s$. Each column represents one complete CDCC solution.

The key step is to find a small set of basis vectors that can represent all the snapshots (and, by extension, solutions at nearby parameter values). This is accomplished using the singular value decomposition (SVD), computed via LAPACK~\cite{LAPACK}.

The SVD factorizes the snapshot matrix as:
\begin{equation}
    \mathbb{C}_{\text{snap}}^J = \mathbf{X}^J \boldsymbol{\Sigma}^J (\mathbf{W}^J)^H,
    \label{eq:svd}
\end{equation}
Here $\mathbf{X}^J$ is an $N_{\text{tot}} \times N_s$ matrix whose columns $\{\mathbf{x}_1^J, \mathbf{x}_2^J, \ldots\}$ are orthonormal vectors called the \textit{left singular vectors} or \textit{POD modes}, forming an optimal basis for representing the snapshots. The diagonal matrix $\boldsymbol{\Sigma}^J$ contains the \textit{singular values} $\sigma_1 \geq \sigma_2 \geq \cdots \geq 0$ in decreasing order, where $\sigma_i$ measures the importance of the $i$-th basis vector. The matrix $\mathbf{W}^J$ contains the \textit{right singular vectors} describing how each snapshot is represented in the new basis, and $H$ denotes the conjugate transpose.

The crucial property of SVD is that it provides the \textit{optimal} low-rank approximation in the Frobenius norm. If we keep only the first $n_b$ basis vectors (corresponding to the $n_b$ largest singular values), the resulting approximation minimizes the total squared error over all snapshots. No other choice of $n_b$ basis vectors can do better.

To decide how many basis vectors to retain, I use an energy criterion based on the singular values. The ``energy'' captured by the first $n_b$ modes is defined as:
\begin{equation}
    E_{n_b} = \frac{\sum_{i=1}^{n_b} \sigma_i^2}{\sum_{i=1}^{N_s} \sigma_i^2}.
\end{equation}
This ratio represents the fraction of the total variance in the snapshots that is explained by the first $n_b$ basis vectors. I retain basis vectors until this ratio exceeds a threshold:
\begin{equation}
    E_{n_b} > 1 - \epsilon_{\text{tol}},
\end{equation}
where $\epsilon_{\text{tol}}$ is a small tolerance, typically $10^{-6}$ or smaller. This ensures that the neglected modes contribute less than $\epsilon_{\text{tol}}$ to the total variance.

In practice, the singular values often decay rapidly, the first few capture most of the variation, while the rest are nearly zero. This rapid decay reflects the smoothness of the parameter dependence: nearby parameters give similar solutions, so the solutions are highly correlated. For the CDCC problem considered here, typically $n_b = 5$--$50$ basis vectors suffice to capture more than $1 - \epsilon_{\text{tol}}$ of the variance.

The retained basis vectors form the \textit{reduced basis} (subscript $r$ for ``reduced''):
\begin{equation}
    \mathbf{X}_r^J = \begin{pmatrix} \mathbf{x}_1^J & \mathbf{x}_2^J & \cdots & \mathbf{x}_{n_b}^J \end{pmatrix}.
\end{equation}

Any solution can now be approximated as a linear combination of the reduced basis vectors:
\begin{equation}
    \mathbf{c}^J(\boldsymbol{\theta}) \approx \mathbf{X}_r^J \boldsymbol{\alpha}^J(\boldsymbol{\theta}) = \sum_{i=1}^{n_b} \alpha_i^J(\boldsymbol{\theta}) \, \mathbf{x}_i^J,
    \label{eq:reduced_approx}
\end{equation}
where $\boldsymbol{\alpha}^J(\boldsymbol{\theta}) = (\alpha_1^J, \alpha_2^J, \ldots, \alpha_{n_b}^J)^T$ is a small vector of $n_b$ \textit{reduced coefficients}. The emulator's task is to determine these $n_b$ coefficients for any new parameter value $\boldsymbol{\theta}$, without solving the full $N_{\text{tot}}$-dimensional system.

Since I construct a separate reduced basis for each partial wave $J$ (because different $J$ values have different numbers of coupled channels and different solution structures), the complete emulator consists of a collection of $J$-dependent reduced bases $\{\mathbf{X}_r^J\}_{J=0}^{J_{\max}}$.

\subsection{Prediction via Galerkin projection}
\label{sec:prediction}
\begin{figure*}[htb]
    \centering
    \includegraphics[width=\textwidth]{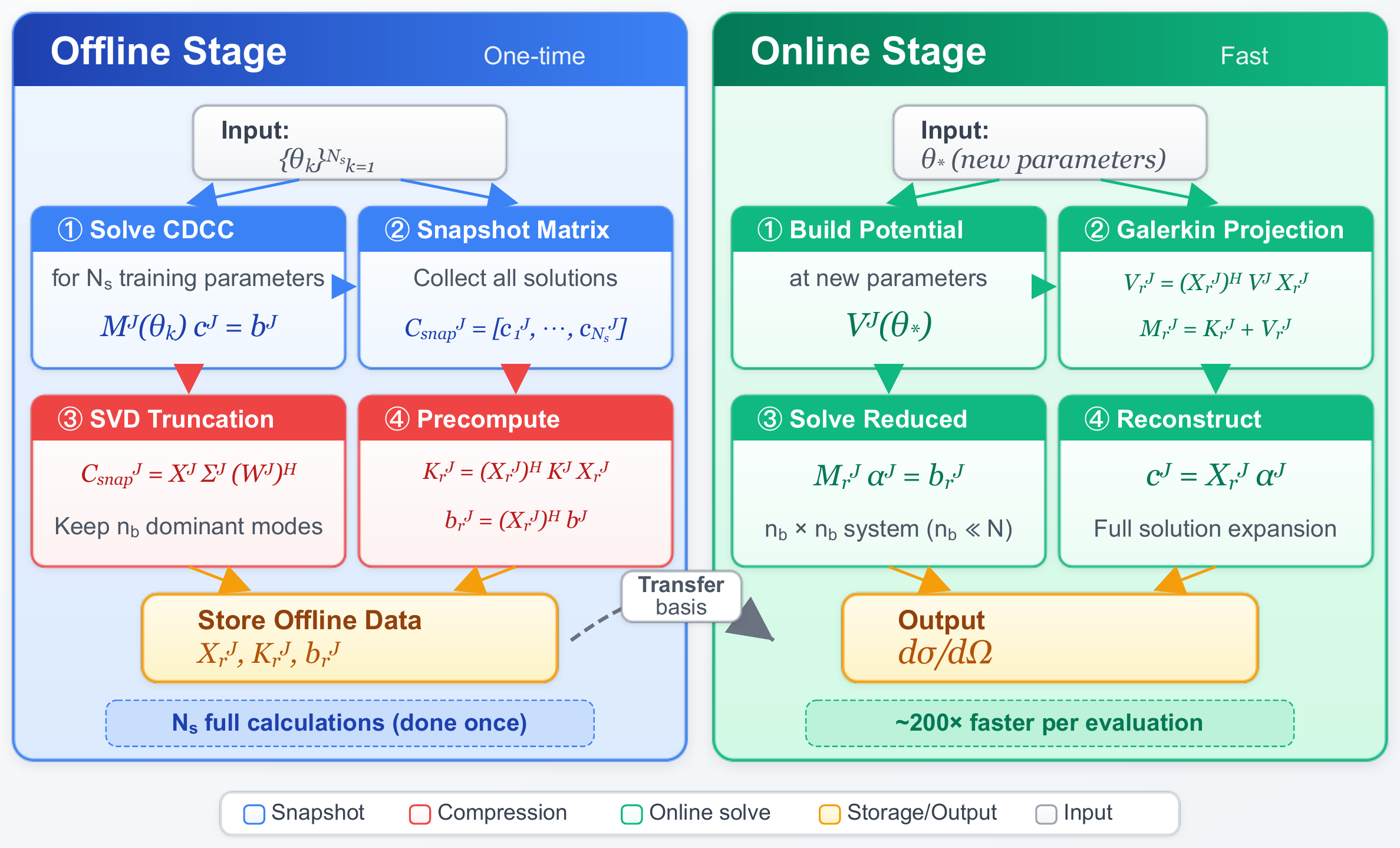}
    \caption{Schematic of the reduced basis emulator workflow. The offline stage (left, blue) generates snapshots by solving the full CDCC problem at $N_s$ sampled parameter values, constructs a joint snapshot matrix for all channels, extracts the reduced basis via SVD truncation, and precomputes parameter-independent matrices. The online stage (right, green) rapidly evaluates new parameter points by building the potential matrix, performing Galerkin projection onto the reduced basis, solving the small $n_b \times n_b$ linear system, and reconstructing the full solution to obtain the differential cross section $d\sigma/d\Omega$.}
    \label{fig:workflow}
\end{figure*}

Once the reduced basis is built, I use the Galerkin projection method to predict the reduced coefficients $\boldsymbol{\alpha}^J$ for a new parameter value $\boldsymbol{\theta}_*$. This approach belongs to the same family of reduced basis methods as the eigenvector continuation technique~\cite{Frame2018,Konig2020}, which has been successfully applied to bound-state problems in nuclear physics. Here I adapt this strategy to the scattering problem formulated as a linear system.

The idea is to substitute the reduced representation~(\ref{eq:reduced_approx}) into the original linear system~(\ref{eq:linear_system}) and project the equation onto the reduced basis. Starting from:
\begin{equation}
    \mathbf{M}^J(\boldsymbol{\theta}_*) \mathbf{c}^J = \mathbf{b}^J,
\end{equation}
I substitute $\mathbf{c}^J \approx \mathbf{X}_r^J \boldsymbol{\alpha}^J$ to get:
\begin{equation}
    \mathbf{M}^J(\boldsymbol{\theta}_*) \mathbf{X}_r^J \boldsymbol{\alpha}^J \approx \mathbf{b}^J.
\end{equation}
Since the approximation cannot satisfy all $N_{\text{tot}}$ equations exactly with only $n_b$ unknowns, I seek a solution that makes the residual orthogonal to the reduced basis. Multiplying both sides from the left by $(\mathbf{X}_r^J)^H$ yields:
\begin{equation}
    \underbrace{(\mathbf{X}_r^J)^H \mathbf{M}^J(\boldsymbol{\theta}_*) \mathbf{X}_r^J}_{\mathbf{M}_r^J} \boldsymbol{\alpha}^J = \underbrace{(\mathbf{X}_r^J)^H \mathbf{b}^J}_{\mathbf{b}_r^J}.
    \label{eq:reduced_system}
\end{equation}
This procedure, called Galerkin projection, yields a reduced linear system of dimension $n_b \times n_b$.

The reduced system~(\ref{eq:reduced_system}) is much smaller than the original (e.g., $20 \times 20$ vs.\ $5000 \times 5000$), but it still incorporates the physics at the new parameter value $\boldsymbol{\theta}_*$ through the matrix $\mathbf{M}^J(\boldsymbol{\theta}_*)$. This ensures high accuracy because the method solves the actual (projected) Schr\"odinger equation at the target parameters.

The computational efficiency comes from decomposing the system matrix as $\mathbf{M}^J = \mathbf{K}^J + \mathbf{V}^J(\boldsymbol{\theta})$, where $\mathbf{K}^J$ contains parameter-independent terms (kinetic energy, centrifugal barrier, boundary conditions) and $\mathbf{V}^J(\boldsymbol{\theta})$ contains the parameter-dependent coupling potentials. During training, I precompute and store:
\begin{equation}
    \mathbf{K}_r^J = (\mathbf{X}_r^J)^H \mathbf{K}^J \mathbf{X}_r^J, \quad \mathbf{b}_r^J = (\mathbf{X}_r^J)^H \mathbf{b}^J.
\end{equation}
At prediction time, only the potential contribution needs to be computed:
\begin{equation}
    \mathbf{M}_r^J(\boldsymbol{\theta}_*) = \mathbf{K}_r^J + (\mathbf{X}_r^J)^H \mathbf{V}^J(\boldsymbol{\theta}_*) \mathbf{X}_r^J.
\end{equation}
Solving the small $n_b \times n_b$ system is then nearly instantaneous.

\subsection{Computational implementation}
\label{sec:implementation}

The computational efficiency of the Galerkin prediction depends critically on how the reduced matrix $\mathbf{M}_r^J$ is assembled. The dominant cost is computing the potential contribution $\mathbf{V}_r^J = (\mathbf{X}_r^J)^H \mathbf{V}^J(\boldsymbol{\theta}_*) \mathbf{X}_r^J$.

I exploit the block structure of the potential matrix $\mathbf{V}^J$. At each radial mesh point $R_i$, the coupling potential has the form of an $N_{\text{ch}} \times N_{\text{ch}}$ matrix $\mathbf{V}_{ij}^J = V_{\alpha\alpha'}(R_i) \delta_{ij}$, which couples all channels at the same radial point. This allows the computation to be organized as:
\begin{equation}
    \mathbf{V}_r^J = \sum_{i=1}^{N-1} \bigl[(\mathbf{X}_r^J)_{(i)}\bigr]^H \mathbf{V}^J(R_i) (\mathbf{X}_r^J)_{(i)},
\end{equation}
where $(\mathbf{X}_r^J)_{(i)}$ denotes the submatrix of $\mathbf{X}_r^J$ corresponding to radial point $R_i$. The sum runs to $N-1$ because the boundary condition at $R_N = R_{\max}$ is handled separately in $\mathbf{K}^J$.

For each radial point $R_i$, the computation involves extracting the $N_{\text{ch}} \times n_b$ submatrix of basis vectors, computing $\mathbf{W}_i = \mathbf{V}(R_i) \mathbf{X}_{(i)}$ using BLAS \texttt{ZGEMM}, and accumulating $\mathbf{V}_r^J \mathrel{+}= (\mathbf{X}_{(i)})^H \mathbf{W}_i$ using another \texttt{ZGEMM} call.

The final reduced matrix is then $\mathbf{M}_r^J = \mathbf{K}_r^J + \mathbf{V}_r^J$, and the $n_b \times n_b$ linear system is solved using LAPACK \texttt{ZGESV}. The reduced source vector $\mathbf{b}_r^J$ is precomputed during training (since $\mathbf{b}^J$ is parameter-independent), and the solution reconstruction $\mathbf{c}^J = \mathbf{X}_r^J \boldsymbol{\alpha}^J$ is computed using BLAS \texttt{ZGEMV}.

This implementation achieves the reported speedups by leveraging highly optimized BLAS/LAPACK libraries (OpenBLAS or Intel MKL), which provide efficient matrix operations with automatic multi-threading.

\subsection{Practical workflow}
\label{sec:workflow}

Figure~\ref{fig:workflow} summarizes the complete emulator workflow, which consists of two stages. The offline stage (left panel) is performed once: $N_s$ parameter samples are generated using Latin hypercube sampling, the full CDCC problem is solved for each sample, and the SVD extracts the optimal reduced basis. The precomputed quantities $\mathbf{X}_r^J$, $\mathbf{K}_r^J$, and $\mathbf{b}_r^J$ are stored for use in the online stage.

The online stage (right panel) is repeated many times during applications such as Bayesian inference. For each new parameter value $\boldsymbol{\theta}_*$, the potential matrix is constructed, the reduced matrix $\mathbf{M}_r^J$ is assembled via Galerkin projection, the small $n_b \times n_b$ linear system is solved, and the full solution is reconstructed to compute observables.

The offline stage requires $N_s$ full CDCC calculations per partial wave, which can be parallelized across parameter samples. Once the emulator is built, each online evaluation is approximately 200 times faster than a full calculation, enabling uncertainty quantification studies that would otherwise be computationally prohibitive.

\section{Results}
\label{sec:results}

I demonstrate the emulator using deuteron elastic scattering on $^{58}$Ni at $E_{\text{lab}} = 21.6$ MeV. The deuteron is described as a proton-neutron system bound by a Gaussian potential~\cite{Austern1987}, and its continuum is discretized into bins for $s$-, $p$-, and $d$-waves with excitation energies up to 12 MeV. For simplicity, nucleon spins are neglected. 

The fragment-target interactions are described by Woods-Saxon optical potentials:
\begin{align}
    U_{fA}(R) = &-V_v h(R; r_v, a_v) - i W_v h(R; r_w, a_w) \notag \\
    &- i W_d \frac{d}{dR} h(R; r_d, a_d) + V_C(R),
\end{align}
where $f \in \{p, n\}$ denotes the fragment (proton or neutron), $h(R; r, a) = [1 + \exp((R - rA_A^{1/3})/a)]^{-1}$ is the Woods-Saxon form factor with $A_A$ being the target mass number, and $V_C$ is the Coulomb potential.

The calculation uses $J_{\max} = 30$ partial waves and $N = 180$ Lagrange mesh points with $R_{\max} = 100$ fm. This results in approximately 37 coupled channels for the largest $J$ values, giving a system dimension of $N_{\text{tot}} \approx 6660$ per partial wave.

I construct an emulator for 18 optical potential parameters: 9 for the proton-target interaction and 9 for the neutron-target interaction. The parameters and their ranges are listed in Table~\ref{tab:params}.

\begin{table}[htb]
\caption{Optical potential parameters and their ranges used for emulator training.}
\label{tab:params}
\begin{ruledtabular}
\begin{tabular}{lcccc}
Parameter & \multicolumn{2}{c}{$p$-$^{58}$Ni} & \multicolumn{2}{c}{$n$-$^{58}$Ni} \\
          & Min & Max & Min & Max \\
\hline
$V_v$ (MeV) & 40.0 & 65.0 & 35.0 & 60.0 \\
$r_v$ (fm) & 1.00 & 1.35 & 1.00 & 1.35 \\
$a_v$ (fm) & 0.55 & 0.95 & 0.55 & 0.95 \\
$W_v$ (MeV) & 0.0 & 15.0 & 0.0 & 15.0 \\
$r_w$ (fm) & 1.10 & 1.50 & 1.00 & 1.45 \\
$a_w$ (fm) & 0.40 & 0.70 & 0.40 & 0.75 \\
$W_d$ (MeV) & 3.0 & 15.0 & 3.0 & 15.0 \\
$r_d$ (fm) & 1.10 & 1.50 & 1.00 & 1.45 \\
$a_d$ (fm) & 0.40 & 0.70 & 0.40 & 0.75 \\
\end{tabular}
\end{ruledtabular}
\end{table}

The emulator is trained using Latin hypercube samples spanning this 18-dimensional parameter space. I compare two training set sizes: $N_{\text{sample}} = 200$ and $N_{\text{sample}} = 400$. The nominal parameter values (center of the ranges) are taken from the Koning-Delaroche global optical potential (KD02)~\cite{Koning2003}.

To evaluate the emulator accuracy, I select 5 test parameter sets that are not included in the training data. Test~1 uses nominal parameters near the center of the parameter space, representing typical phenomenological values. Tests~2 and 3 modify only the proton-target or neutron-target potential respectively, testing sensitivity to individual fragment interactions. Test~4 modifies both potentials simultaneously, probing the emulator's ability to handle correlated parameter changes. Test~5 uses parameters near the edge of the training range with strong volume absorption ($W_v \approx 10$--12~MeV), providing a challenging test at the boundary of the parameter space.

The specific parameter values for all five test cases are listed in Table~\ref{tab:test_params}. The test cases span a significant portion of the parameter space, with real potential depths varying from 38 to 55~MeV and surface absorption strengths from 5 to 12~MeV.

\begin{table*}[htb]
\caption{Optical potential parameters for the five test cases. Parameters are grouped by fragment: proton-target ($p$-$^{58}$Ni) and neutron-target ($n$-$^{58}$Ni). All lengths are in fm and energies in MeV.}
\label{tab:test_params}
\begin{ruledtabular}
\begin{tabular}{cl|ccccccccc}
Test & Fragment & $V_v$ & $r_v$ & $a_v$ & $W_v$ & $r_w$ & $a_w$ & $W_d$ & $r_d$ & $a_d$ \\
\hline
\multirow{2}{*}{1} & $p$-$^{58}$Ni & 53.3 & 1.17 & 0.75 & 0.0 & 1.32 & 0.53 & 7.8 & 1.32 & 0.53 \\
                   & $n$-$^{58}$Ni & 48.5 & 1.17 & 0.75 & 0.0 & 1.26 & 0.58 & 9.2 & 1.26 & 0.58 \\
\hline
\multirow{2}{*}{2} & $p$-$^{58}$Ni & 50.0 & 1.20 & 0.70 & 2.0 & 1.35 & 0.55 & 10.0 & 1.35 & 0.55 \\
                   & $n$-$^{58}$Ni & 48.5 & 1.17 & 0.75 & 0.0 & 1.26 & 0.58 & 9.2 & 1.26 & 0.58 \\
\hline
\multirow{2}{*}{3} & $p$-$^{58}$Ni & 53.3 & 1.17 & 0.75 & 0.0 & 1.32 & 0.53 & 7.8 & 1.32 & 0.53 \\
                   & $n$-$^{58}$Ni & 45.0 & 1.20 & 0.80 & 3.0 & 1.30 & 0.60 & 12.0 & 1.30 & 0.60 \\
\hline
\multirow{2}{*}{4} & $p$-$^{58}$Ni & 55.0 & 1.15 & 0.72 & 1.5 & 1.30 & 0.50 & 9.0 & 1.30 & 0.50 \\
                   & $n$-$^{58}$Ni & 50.0 & 1.15 & 0.72 & 2.0 & 1.25 & 0.55 & 10.5 & 1.25 & 0.55 \\
\hline
\multirow{2}{*}{5} & $p$-$^{58}$Ni & 42.0 & 1.05 & 0.60 & 12.0 & 1.45 & 0.65 & 5.0 & 1.45 & 0.65 \\
                   & $n$-$^{58}$Ni & 38.0 & 1.05 & 0.60 & 10.0 & 1.40 & 0.70 & 5.5 & 1.40 & 0.70 \\
\end{tabular}
\end{ruledtabular}
\end{table*}

For each test case, I compute the full CDCC solution and compare with the emulator predictions.

\begin{figure*}[!htb]
    \centering
    \includegraphics[width=0.95\textwidth]{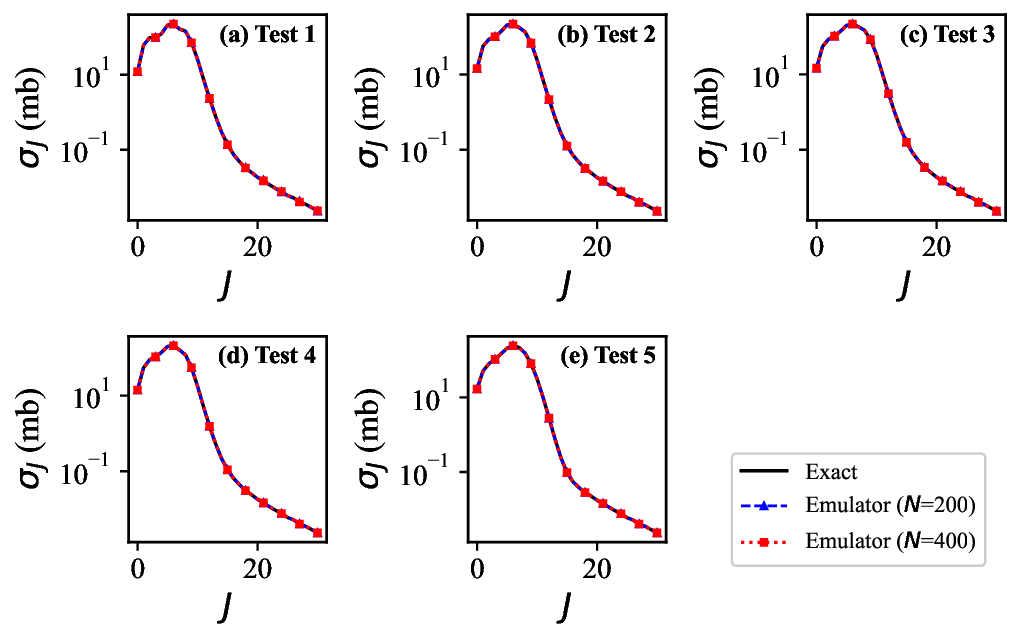}
    \caption{Partial-wave elastic cross sections $\sigma_J$ versus total angular momentum $J$ for five test parameter sets. Black solid lines: exact CDCC calculations. Blue dashed lines with triangles: emulator predictions with $N_{\text{sample}} = 200$. Red dotted lines with squares: emulator predictions with $N_{\text{sample}} = 400$. The three curves overlap almost perfectly.}
    \label{fig:sigma_vs_J_all}
\end{figure*}

Figure~\ref{fig:sigma_vs_J_all} shows the partial-wave elastic cross sections $\sigma_J$ as a function of total angular momentum $J$ for all five test cases. Panels (a)--(e) correspond to Tests~1--5, respectively. In each panel, the black solid lines represent the exact CDCC calculations, the blue dashed lines with triangles show the emulator predictions trained with $N_s = 200$ samples, and the red dotted lines with squares show the emulator predictions trained with $N_s = 400$ samples. The cross sections span several orders of magnitude, peaking around $J = 5$--$7$ and decreasing rapidly for higher $J$ values as the centrifugal barrier suppresses the nuclear interaction. The three curves overlap almost perfectly in all panels, with the emulator predictions agreeing with the exact calculations to within 0.1\% across all partial waves and test cases.

\begin{figure}[!htb]
    \centering
    \includegraphics[width=0.95\columnwidth]{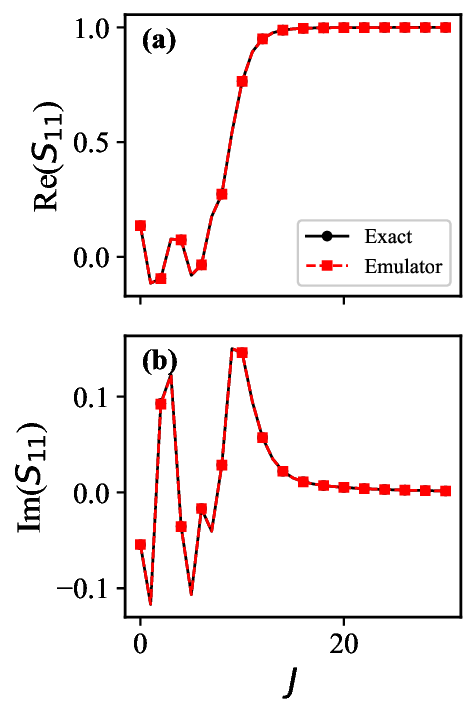}
    \caption{Elastic $S$-matrix element $S_{11}^J$ for Test~1 with the emulator trained using $N_s = 200$ samples. (a) Real part. (b) Imaginary part. Black circles with solid lines: exact CDCC calculation. Red squares with dashed lines: emulator prediction.}
    \label{fig:Smatrix_components}
\end{figure}

The elastic $S$-matrix element $S_{11}^J$, which describes the ground-state-to-ground-state elastic scattering amplitude, provides a more stringent test of the emulator accuracy. Figure~\ref{fig:Smatrix_components} shows the real and imaginary parts of $S_{11}^J$ for Test~1 using the emulator trained with $N_s = 200$ samples. Panel~(a) displays the real part and panel~(b) displays the imaginary part; in both panels, black circles with solid lines represent the exact CDCC calculation, while red squares with dashed lines show the emulator prediction. At low $J$, the $S$-matrix shows strong absorption ($|S_{11}| \ll 1$) and rapid oscillations reflecting the complex nuclear interaction. As $J$ increases, the system becomes more peripheral and $S_{11} \to 1$ as the nuclear interaction becomes negligible. The emulator accurately reproduces both the oscillatory behavior at low $J$ and the smooth approach to unity at high $J$.

\begin{figure*}[!htb]
    \centering
    \includegraphics[width=0.95\textwidth]{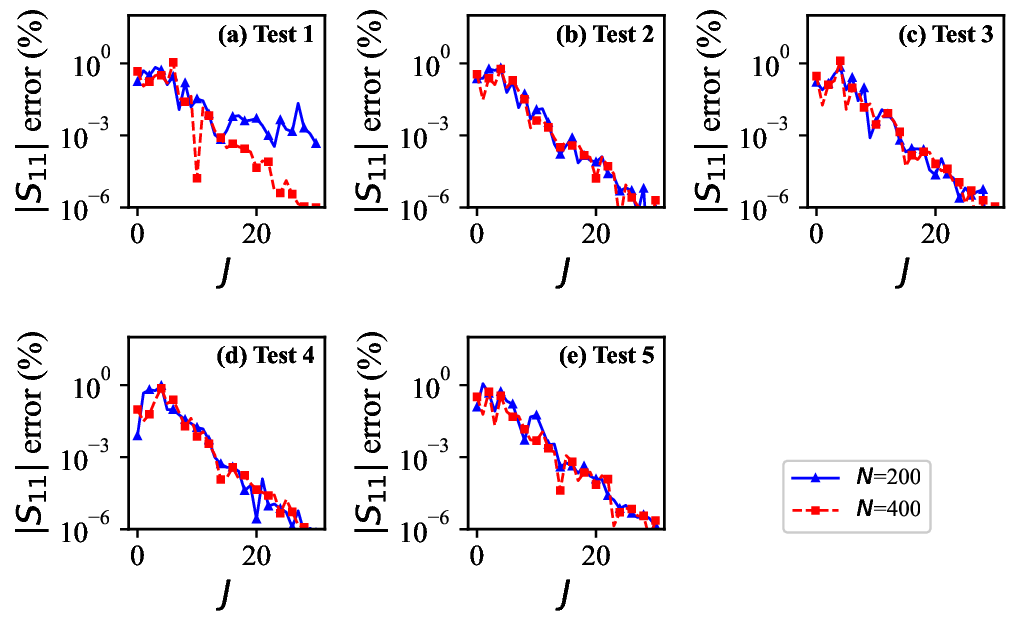}
    \caption{Relative error in the magnitude of the elastic $S$-matrix element $|S_{11}^J|$ versus $J$ for five test cases. Blue solid lines with triangles: $N_{\text{sample}} = 200$. Red dashed lines with squares: $N_{\text{sample}} = 400$.}
    \label{fig:Smatrix_error}
\end{figure*}

Figure~\ref{fig:Smatrix_error} quantifies the relative error in $|S_{11}^J|$ for all five test cases. Panels (a)--(e) correspond to Tests~1--5, respectively. In each panel, the blue solid lines with triangles show the error for the emulator trained with $N_s = 200$ samples, while the red dashed lines with squares show the error for $N_s = 400$ samples. The errors remain below 1\% for all $J$ values and test cases, with most partial waves showing errors in the range $10^{-4}$--$10^{-2}$\%. The two training set sizes yield comparable accuracy, indicating that $N_s = 200$ is sufficient for this 18-dimensional parameter space.

\begin{figure}[!htb]
    \centering
    \includegraphics[width=0.95\columnwidth]{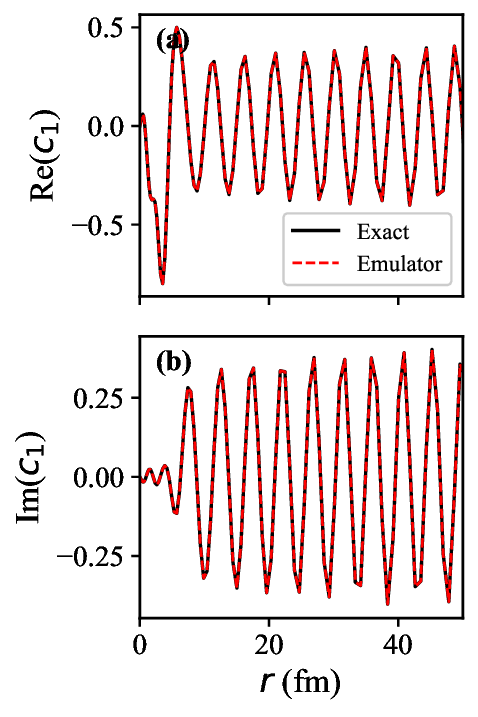}
    \caption{Elastic channel wave function coefficient $c_1(r)$ for $J = 0$ (Test~1) with the emulator trained using $N_s = 200$ samples. (a) Real part. (b) Imaginary part. Black solid lines: exact CDCC calculation. Red dashed lines: emulator prediction.}
    \label{fig:wavefunction}
\end{figure}

The wave function coefficient $c_1(r)$ represents the most detailed test of emulator accuracy, as it contains the full radial dependence of the scattering wave function. Figure~\ref{fig:wavefunction} compares the elastic channel wave function coefficient $c_1(r)$ for $J = 0$ between the exact CDCC calculation and the emulator prediction for Test~1 using $N_s = 200$ training samples. Panel~(a) shows the real part and panel~(b) shows the imaginary part; in both panels, black solid lines represent the exact calculation while red dashed lines show the emulator prediction. The wave function shows characteristic oscillations corresponding to the interference between incoming and scattered waves. The emulator reproduces both the amplitude and phase of these oscillations throughout the entire radial range, with the two curves virtually indistinguishable.

\begin{figure}[!htb]
    \centering
    \includegraphics[width=0.95\columnwidth]{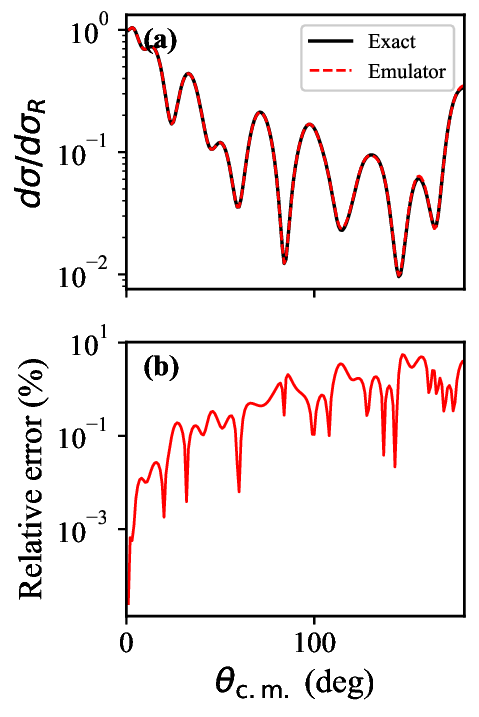}
    \caption{Elastic angular distribution for Test~1 with $N_{\text{sample}} = 200$. (a) Ratio to Rutherford cross section. Black solid line: exact calculation. Red dashed line: emulator prediction. (b) Relative error in the angular distribution.}
    \label{fig:angular}
\end{figure}

The elastic angular distribution provides an integrated test of the emulator, as it combines contributions from all partial waves through the scattering amplitude. Figure~\ref{fig:angular} compares the elastic angular distribution between the exact CDCC calculation and the emulator prediction for Test~1 using $N_s = 200$ training samples. Panel~(a) shows the ratio of the elastic differential cross section to the Rutherford cross section, $d\sigma/d\sigma_R$, as a function of center-of-mass scattering angle; black solid lines represent the exact calculation while red dashed lines show the emulator prediction. The angular distribution exhibits the characteristic Fresnel diffraction pattern with oscillations at forward angles arising from Coulomb-nuclear interference and a smooth falloff at backward angles. Panel~(b) displays the relative error between the emulator and exact calculations.

The relative error [Fig.~\ref{fig:angular}(b)] remains below 0.1\% for most angles, with slightly larger errors near the diffraction minima where the cross section passes through zero. This excellent agreement demonstrates that the emulator correctly captures the delicate interference between Coulomb and nuclear scattering.

Table~\ref{tab:accuracy} summarizes the total elastic cross sections and emulator errors for all five test cases. The exact cross sections range from 1169 to 1341~mb, reflecting the variation in optical potential parameters across the test set. Both the $N_{\text{sample}} = 200$ and $N_{\text{sample}} = 400$ cases use the same reduced basis dimension $n_b = 50$, which explains why increasing the training set size does not show a clear improvement trend. Nevertheless, even with this modest basis size, the emulator errors are consistently below 0.05\%, demonstrating excellent agreement with the exact calculations.

\begin{table}[htb]
\caption{Total elastic cross sections and emulator relative errors for five test parameter sets. Results are shown for $N_{\text{sample}} = 200$ and 400.}
\label{tab:accuracy}
\begin{ruledtabular}
\begin{tabular}{ccccc}
Test & $\sigma_{\text{exact}}$ & \multicolumn{2}{c}{Relative error (\%)} \\
     & (mb) & $N_{\text{sample}} = 200$ & $N_{\text{sample}} = 400$ \\
\hline
1 & 1245.50 & 0.0049 & 0.0245 \\
2 & 1236.12 & 0.0074 & 0.0067 \\
3 & 1277.56 & 0.0156 & 0.0097 \\
4 & 1168.76 & 0.0097 & 0.0198 \\
5 & 1340.93 & 0.0426 & 0.0294 \\
\hline
Average & --- & 0.016 & 0.018 \\
\end{tabular}
\end{ruledtabular}
\end{table}

\begin{figure}[!htb]
    \centering
    \includegraphics[width=0.95\columnwidth]{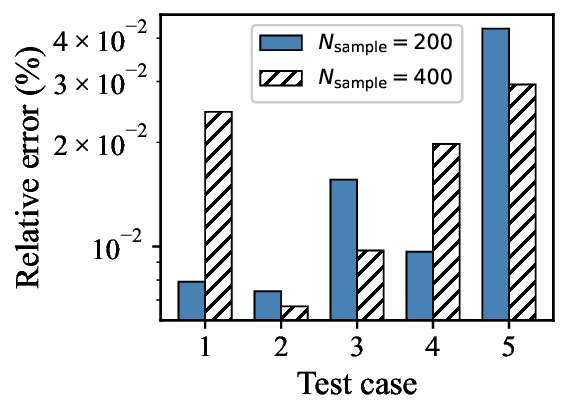}
    \caption{Comparison of total cross section errors for $N_{\text{sample}} = 200$ (solid blue bars) and $N_{\text{sample}} = 400$ (hatched white bars) across five test cases.}
    \label{fig:nsample_comparison}
\end{figure}

Figure~\ref{fig:nsample_comparison} compares the emulator accuracy for two training set sizes: $N_{\text{sample}} = 200$ and 400. Doubling the training set size does not uniformly improve the accuracy. For some test cases (Tests~2 and~3), the larger training set gives slightly better results, while for others (Tests~1 and~4), the smaller training set performs marginally better.

This behavior indicates that both training set sizes have effectively converged to the same level of accuracy. The observed differences (on the order of 0.01--0.02\%) are comparable to numerical precision effects and do not represent a systematic trend. The SVD-based reduced basis construction automatically determines the optimal number of basis vectors based on the tolerance threshold $\epsilon_{\text{tol}} = 10^{-6}$; in this case, approximately 50 basis vectors per partial wave suffice to capture the relevant variation across the 18-dimensional parameter space. Once this number of basis vectors is reached, adding more training samples does not significantly improve accuracy because the basis already spans the relevant subspace.

The key finding is that 200 training samples are sufficient to achieve sub-0.05\% accuracy in an 18-dimensional parameter space. This favorable scaling, requiring only $\approx 10$ samples per parameter dimension rather than the exponentially many samples expected from naive grid-based approaches, demonstrates the efficiency of the reduced basis method for smooth parameter dependences.

Table~\ref{tab:timing} compares the computational cost of emulator predictions with full CDCC calculations. All timings are wall-clock times measured on a server with dual Intel Xeon Gold 6248R processors (3.0 GHz, 48 cores total) using OpenBLAS with multi-threading.

\begin{table}[htb]
\caption{Computational cost comparison between full CDCC calculation and emulator prediction. Times are for a single partial wave $J$.}
\label{tab:timing}
\begin{ruledtabular}
\begin{tabular}{lcc}
Method & Time & Speedup \\
\hline
Full CDCC (direct solve) & 6.5 s & --- \\
Emulator prediction & 30 ms & $\approx 220\times$ \\
\end{tabular}
\end{ruledtabular}
\end{table}

The emulator achieves a speedup of approximately 220$\times$ compared to the full CDCC calculation. This speedup makes it feasible to perform Bayesian inference with $10^5$--$10^6$ likelihood evaluations: what would take months with direct calculations can be completed in hours with the emulator.

The training cost scales linearly with the number of samples and partial waves. For $N_{\text{sample}} = 200$ and 31 partial waves, training requires approximately 11 hours ($6.5~\text{s} \times 31 \times 200 / 3600 \approx 11~\text{hr}$). This one-time cost is easily amortized over subsequent predictions: a complete evaluation at one parameter point requires 31 partial waves, so full CDCC takes $\approx 200$~s while the emulator takes $\approx 1$~s. After approximately 200 emulator evaluations, the total time breaks even with direct calculations, and for the $10^5$--$10^6$ evaluations typical of Bayesian inference, the training cost becomes negligible.

\section{Discussion and Conclusions}
\label{sec:conclusion}

I have developed a reduced basis emulator for CDCC calculations that achieves substantial speedups while maintaining high accuracy. The approach features per-$J$ basis construction, where separate reduced bases for each partial wave naturally accommodate the varying channel structure. Predictions are performed by solving a reduced linear system obtained via Galerkin projection, which incorporates the physics at the target parameters and ensures high accuracy. The emulator successfully handles 18 optical potential parameters, demonstrating scalability to realistic uncertainty quantification scenarios. The accuracy achieved (sub-percent errors in cross sections) is sufficient for most nuclear physics applications.

Several extensions of this work are possible. The emulator enables efficient Markov chain Monte Carlo sampling for Bayesian parameter inference, extracting optical potential parameters from experimental data with full uncertainty quantification. Breakup reactions of halo nuclei, which require large model spaces, would particularly benefit from emulation. Extending the emulator to cover a range of beam energies would further increase its utility for systematic studies.

Recently, reduced basis emulators have been developed for coupled-channel calculations in nuclear reactions. Catacora-Rios \textit{et al.}~\cite{CatacoraRios2025} constructed a coupled-channel emulator for inelastic scattering using channel-wise basis functions combined with the empirical interpolation method (EIM) for non-affine potentials. Their approach achieves speedups of $\approx 30\times$ for systems with 2--5 coupled channels, emulating 10 parameters (one deformation length plus nine Woods-Saxon parameters). However, constructing separate bases for each channel leads to computational scaling as $O(N_c^4)$ with the number of channels $N_c$, which limits applications to problems with few channels.

Liao \textit{et al.}~\cite{Liao2025} developed an eigenvector continuation emulator for sub-barrier fusion reactions, combining the discrete basis method with the Kohn variational principle. Their emulator successfully extracts deformation parameters ($\beta_2$, $\beta_4$) from fusion cross sections with speedups of 200--400$\times$. However, their parameter space is restricted to only 2--4 deformation parameters, with the optical potential parameters held fixed. The channel structure in fusion calculations (rotational bands with $N_{\text{ch}} \lesssim 7$) is also considerably simpler than in CDCC.

Both of these approaches operate in relatively low-dimensional parameter spaces, which limits their applicability to comprehensive uncertainty quantification studies where all relevant model parameters, including those of fragment-target optical potentials, must be varied simultaneously.

The present work demonstrates CDCC emulation with $N_{\text{ch}} \approx 30$--50 coupled channels arising from continuum discretization, while simultaneously exploring a high-dimensional parameter space of 18 optical potential parameters. Importantly, the method scales favorably to even larger channel numbers in principle. By constructing a single reduced basis for the entire coefficient vector across all channels, the basis dimension $n_b$ is determined by the singular value decay rather than the channel number. This favorable scaling, combined with the ability to handle many parameters, makes the present emulator well-suited for Bayesian inference where the full optical potential parameter space must be explored. The direct computation of potential matrices at prediction time, rather than using EIM, simplifies implementation without sacrificing efficiency for the channel numbers typical in CDCC.

For halo nuclei such as $^{11}$Be, where adequate discretization of the extended continuum may require $N_{\text{ch}} \approx 1000$ coupled channels, the computational scaling of the online stage warrants consideration. The dominant cost lies in constructing the reduced potential matrix $\mathbf{V}_r^J$, which requires $O(N \times N_{\text{ch}}^2 \times n_b)$ operations for the two matrix multiplications at each radial mesh point. This quadratic scaling with $N_{\text{ch}}$ is more favorable than the $O(N^3 \times N_{\text{ch}}^3)$ scaling of full CDCC solutions. Crucially, the reduced basis dimension $n_b$ is governed by the smoothness of parameter dependence rather than the channel count; preliminary analysis suggests that $n_b$ remains modest ($\lesssim 100$) even as $N_{\text{ch}}$ increases substantially, provided the solution manifold retains its low-dimensional structure. The offline training cost, dominated by full CDCC solves for building the snapshot matrix, will increase with $N_{\text{ch}}$, but this one-time investment is amortized over subsequent predictions. Systematic investigation of the emulator performance for halo systems represents an important direction for future work.

In conclusion, reduced basis emulation provides a powerful tool for accelerating CDCC calculations by orders of magnitude. This development opens the door to rigorous uncertainty quantification and Bayesian inference for nuclear reaction theory, bringing these calculations into the era of precision nuclear physics.

\begin{acknowledgments}
This work was supported by the National Natural Science Foundation of China (Grant Nos.~12475132 and 12535009) and the Fundamental Research Funds for the Central Universities.
\end{acknowledgments}

\bibliography{references}

\end{document}